\documentclass[twocolumn,twoside,slac_two]{revtex4}
\usepackage{graphicx}
\usepackage{fancyhdr}
\usepackage{amssymb}
\usepackage{natbib}

\newcommand{\fermi}{\emph{Fermi}~}

\def\swift{\emph{Swift}~}

\begin{document}

\title{Monitoring the Gamma-ray Sky through 4.5 Years of Fermi LAT\\ Flare Advocate Service}

%

\author{Stefano Ciprini}
\affiliation{ASI Science Data Center, Frascati, Roma, Italy}%
\affiliation{INAF Observatory of Rome, Monte Porzio Catone, Roma, Italy\\}%


\author{David J. Thompson}
\affiliation{NASA Goddard Space Flight Center, Greenbelt, MD 20771, USA\\ \\(on behalf of the Fermi LAT collaboration)}
%
%
%
\begin{abstract}
The \fermi Flare Advocate (also known as Gamma-ray Sky Watcher, FA-GSW) service provides for a quick look and review of the $\gamma$-ray sky observed daily by the \fermi Large Area Telescope (LAT). The FA-GSW service  provides alerts and communicates to the external scientific community potentially new $\gamma$-ray sources, interesting transients and source flares. A weekly digest containing the highlights about the variable LAT $\gamma$-ray sky is published in the web (``Fermi Sky Blog''). Other news items are occasionally posted through the \fermi  multiwavelength mailing list, Astronomer's Telegrams (ATels) and Gamma-ray Coordination Network notes (GCNs). From July 2008 to January 2013 about 230 ATels and some GCNs have been published by the \fermi LAT Collaboration, more than 40 target of opportunity observing programs have been triggered by the LAT Collaboration and performed though the \swift satellite, and individual observing alerts have been addressed to ground-based Cherenkov telescopes. This is helping the \fermi mission to catch opportunities offered by the variable high-energy sky, increasing the rate of simultaneous multifrequency observations and the level of international scientific cooperation.
\end{abstract}

\maketitle

\thispagestyle{fancy}



\section{Introduction}

The Large Area Telescope (LAT), on board the \fermi {\it Gamma-ray Space Telescope}, is a pair-conversion $\gamma$-ray detector designed to distinguish $\gamma$ rays in the energy range 20 MeV to more than 300 GeV from the intense background of energetic charged particles found in the 565 km altitude orbit of the satellite. For each $\gamma$-ray photon event the LAT measures its arrival time, direction, and energy. The effective collecting area is $\sim 6500$ cm$^2$ at 1 GeV (for the Pass 7 event class selection), the field of view is quite large ($>2$ sr),
and the observing efficiency is very high, limited primarily by interruptions of data taking during passage of Fermi through the South Atlantic Anomaly ($\sim 15\%$) and trigger dead time fraction ($\sim 9\%$). The single-photon angular resolution is strongly dependent on energy;
the 68\% containment radius is about 0.8$^\circ $ at 1 GeV (averaged over the acceptance of the LAT) and varies with energy approximately as E$^{-0.8}$, with an asymptotic value  $\sim 0.2^\circ $ at high energies.

%
\begin{figure}[t!!!]
{\resizebox{7.5cm}{!}{\rotatebox[]{0}{\includegraphics{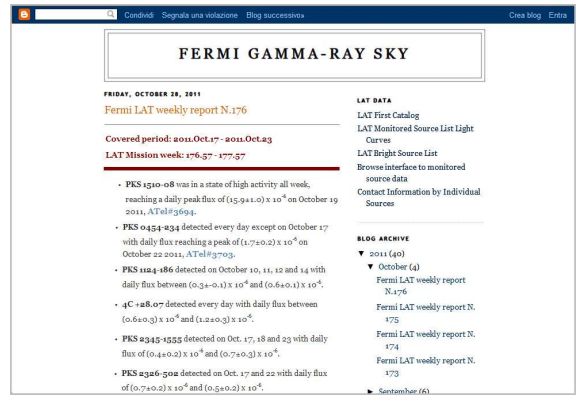}}}}
 \caption{Weekly digests containing the highlights of the $\gamma$-ray sky seen by \fermi LAT are posted every week on the Fermi Sky Blog ($\texttt{http://fermisky.blogspot.com}$).}
   \label{fig:fermiskyblog}
\vspace*{-0.5cm}
\end{figure}

The LAT consists of a tracker, a calorimeter, and an anti-coincidence system. The tracking section of the LAT has 36 layers of silicon strip detectors to record the tracks of charged particles, interleaved with 16 layers of tungsten foil (12 thin layers, 0.03 radiation length, at the top or ``Front'' of the instrument, followed by 4 thick layers, 0.18 radiation length,
in the ``Back'' section) to promote $\gamma$-ray pair conversion. Beneath the tracker is a calorimeter comprised of an 8-layer array of CsI crystals (1.08 radiation length per layer) to determine the $\gamma$-ray photon energy. The tracker is surrounded by a segmented charged-particle anticoincidence detector (plastic scintillators with photomultiplier tubes) to reject cosmic-ray background events. More information about the LAT and its performance is presented in \cite{atwood09}, and the in-flight calibration of the LAT is described in \citet{latcalibration09} and \citet{latcalibration12}.

%
\begin{figure*}[t!!!]
\vspace*{-0.3cm}
{\resizebox{0.80\textwidth}{!}{\rotatebox[]{90}{\includegraphics{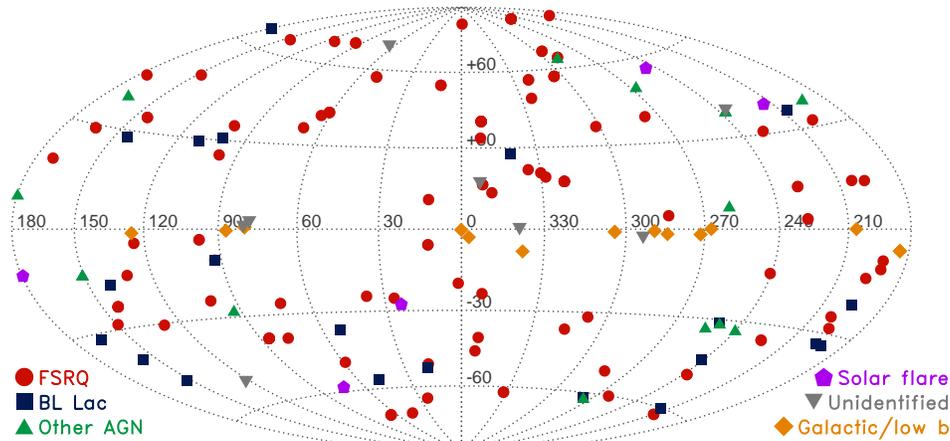}}}}
\vspace*{-4.2cm}
 \caption{All-sky distribution, in Galactic coordinates, of the new $\gamma$-ray sources, flaring $\gamma$-ray blazars/AGN, low Galactic latitude transients and other candidate $\gamma$-ray sources found by \fermi LAT from 2008 July 24 to 2013 January 30 that were subject of Astronomer's Telegrams (from ATel\#1628 to ATel\#4770, see \texttt{www.asdc.asi.it/feratel/}). These LAT sources are represented with different symbols for different types.}
   \label{fig:ATELsourcessky}
\end{figure*}
%

%
\begin{figure}[t!!!]
\begin{center}
\vspace*{-0.5cm}
\resizebox{7.8cm}{!}{\rotatebox[]{0}{\includegraphics{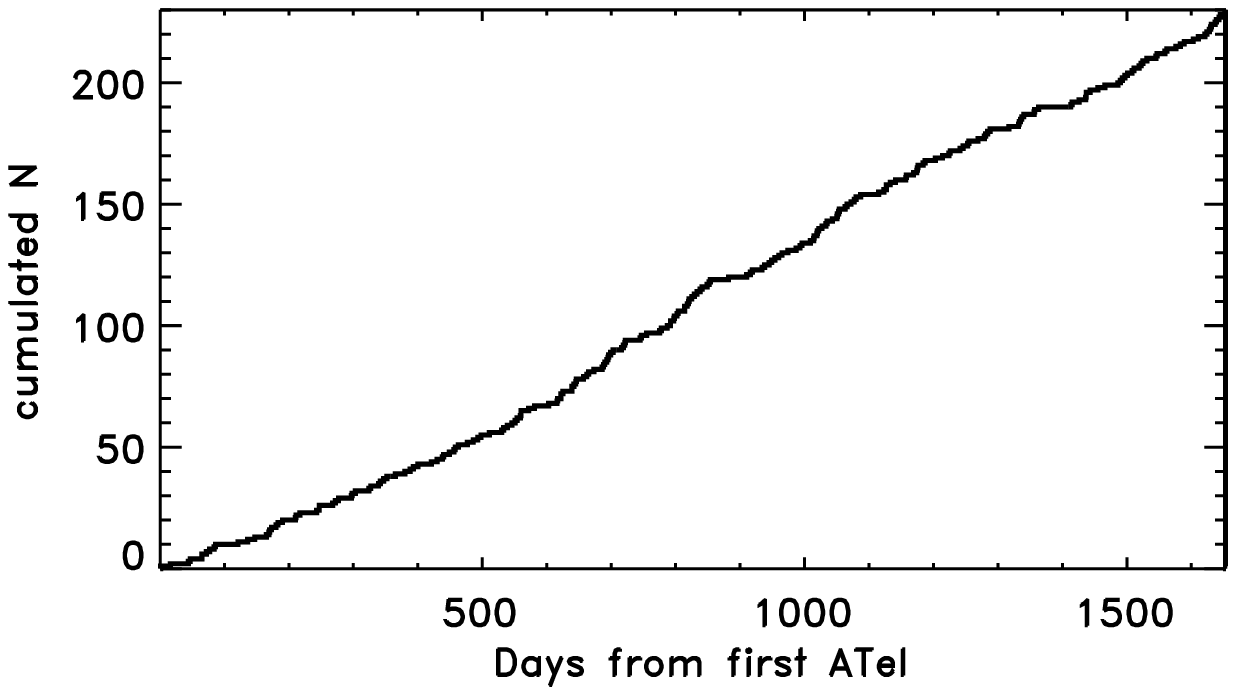}}}\\
\vspace*{-0.5cm}
\resizebox{7.8cm}{!}{\rotatebox[]{0}{\includegraphics{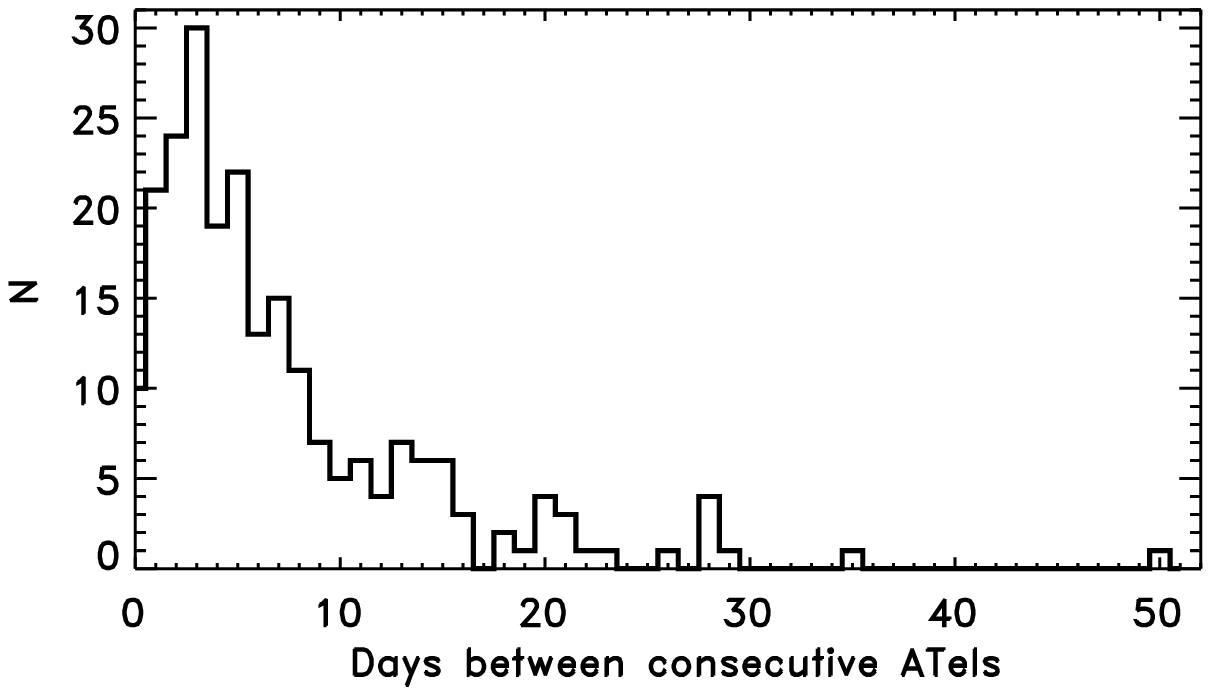}}}
\vspace*{-0.4cm}
\caption{Time distributions of the 229 Astronomer's Telegrams (ATels) published on behalf of the \fermi LAT Collaboration from 2008 July 24 (ATel\#1628) to 2013 January 30 (ATel\#4770), i.e. 1652 days ($\sim 4.5 years$).
}
   \label{fig:ATELstastistics}
\end{center}
\end{figure}
%

Owing to the nearly continuous all-sky survey observing mode, large field of view, good angular resolution and high sensitivity the LAT is an optimal hunter for high-energy flares, transients, and newly appearing $\gamma$-ray sources, providing also an unprecedented monitor of variable $\gamma$-ray sources. This all-sky monitor is complemented by the \fermi Flare Advocate (a.k.a. Gamma-ray Sky Watcher, FA-GSW) service \citep[see, e.g., ][]{ciprini12a,ciprini12b,bastieri11}, belonging to the LAT Instrument Science Operations. The FA-GSW activity is performed with continuity, day-by day all year, with weekly shifts by members and affiliated scientists of the LAT Collaboration.

%
\section{The  Fermi Flare Advocate and Gamma-ray Sky Watcher Service}
%

The FA-GSW service was created to supply a quick-look inspection and daily review of the $\gamma$-ray sky observed by \fermi LAT. The FA-GSW on shift points out basic facts and information about the high-energy sky of potential interest for the LAT science groups, reviewing the all-sky photon count maps collected on short time intervals (one-day and six-hour count maps), reviewing preliminary automated source detections and checking the results of the quick-look science analysis pipeline. For particularly interesting sources a maximum likelihood spectral fit and check of detection significance, flux, spectral photon index, exposure map, localization and multifrequency counterpart associations is carried out.
A concise daily report is compiled day by day by FA-GSWs internally to the LAT Collaboration. About 60 shifters belonging to the LAT Collaboration have served as least one time as FA-GSW.

%
\begin{figure*}[t!!!]
{\resizebox{0.98\textwidth}{!}{\rotatebox[]{0}{\includegraphics{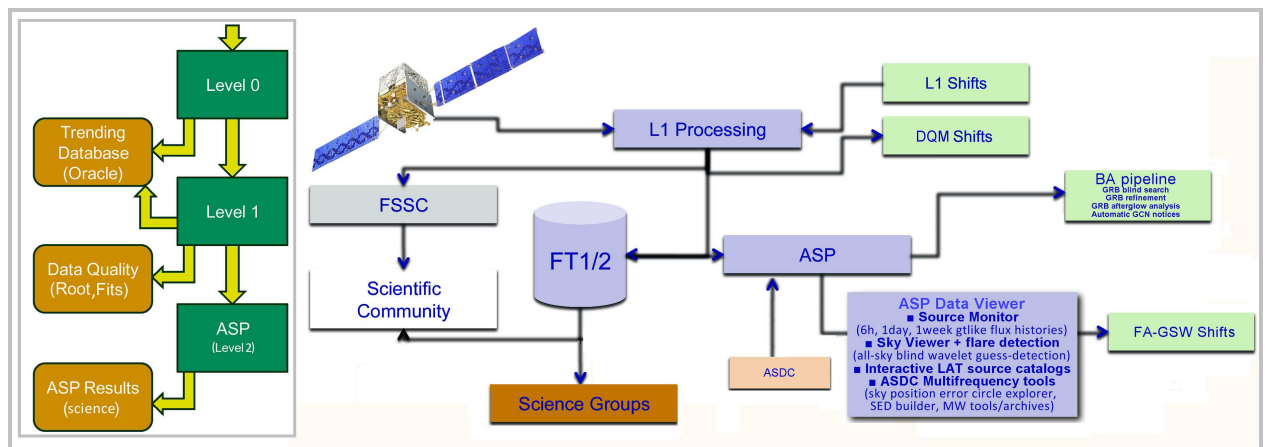}}}}
\vspace*{-0.3 cm}
\caption{\textit{Left inset panel:} diagram of the main general blocks of the \fermi LAT data processing flow at ISOC (SLAC, Stanford University). The astrophysical science data (FT1/FT2 fits files) are sent to the Fermi Science Support Center (FSSC) at the NASA Goddard Space Flight Center
for public distribution to the scientific community.  \textit{Main panel:} block diagram of \fermi LAT data processing with final steps of duty services covered by members of the LAT collaboration in evidence (L1, Data Quality Monitor, DQM, Flare Advocate Gamma-ray Sky Watcher FA-GSW, weekly shifts). In particular the Automated Science Processing (ASP, Level 2 processing) is outlined with the links to the Gamma-ray Burst (GRB) automatic analysis pipeline and to the FA-GSW service.}
   \label{fig:schemeblocks.eps}
\end{figure*}
%

Summaries about significant \fermi LAT source detections, new candidate sources and transients, increasing brightness trends and flares from known sources observed in one-day and six-hour time intervals, are published through regular weekly reports in the ``Fermi Sky Blog''\footnote{ \texttt{
http://fermisky.blogspot.com}} (Figure \ref{fig:fermiskyblog}),  Astronomer's Telegrams (ATels)\footnote{ \texttt{http://www.astronomerstelegram.org}, and\\
\texttt{http://www-glast.stanford.edu/cgi-bin/pub\_rapid} } (Figures \ref{fig:ATELsourcessky} and \ref{fig:ATELstastistics}) , and GCNs for blazar flares and transients\footnote{\texttt{http://gcn.gsfc.nasa.gov/gcn/gcn3\_archive.html}}.  Figure \ref{fig:ATELsourcessky} is the all-sky map of the new $\gamma$-ray sources, flares and transients found by \fermi LAT and the subject of published ATels, in the period between 2008 July 24, and 2013 January 30\footnote{An interactive list of ATel sources can be seen at\\  \texttt{http://www.asdc.asi.it/feratel/} }. The figure shows that most of the flaring sources are blazars as found in the  \fermi LAT source catalogs \citep{2FGL,2LAC,latvariability}. Figure \ref{fig:ATELstastistics} reports the cumulative distribution and time separation of the 229 ATels
published on behalf of the \fermi LAT Collaboration in this period.

Results are also communicated to the multiwavelength (MW)  community using the \fermi MW mailing-list\footnote{Sign up for the ``gammamw'' mailing list at\\   \texttt{https://lists.nasa.gov/mailman/listinfo/gammamw}}.
The FA-GSW service identifies first seeds for MW coordinated observations and studies. The LAT MW Coordinating Group page\footnote{\texttt{https://confluence.slac.stanford.edu/x/YQw}} and the FSSC MW Support page\footnote{\texttt{http://fermi.gsfc.nasa.gov/ssc/observations/multi/}} are other resources with information and points of contacts to coordinate and support  MW observations involving the \fermi LAT. The FA-GSW service and the MW Coordinating Group allow the \fermi mission to promote and increase the rate of multifrequency international collaborations and observing campaigns, maximizing the scientific return.


The FA-GSW routine is based on the LAT data analysis of Level 2 (L2) at the \fermi LAT Instrument Science Operation Center (ISOC) of SLAC, Stanford, \citep{cameron07,chiang07}. L2 processing (instrument monitoring pipeline, background monitoring, quick look science analysis) is triggered by the first availability of Level 1 (L1) processed data and is performed on several time intervals (6 hours, 1 day, 1 week). This Automated Science Processing (ASP, Figure \ref{fig:schemeblocks.eps}) forms the primary input for the FA-GSWs.

The ASP analysis pipeline running on the final astrophysical science data (photon event files FT1, and spacecraft data files FT2 fits files) is composed of several scientific tasks \citep{chiang07}: 1) automatic analysis of $\gamma$-ray bursts (impulsive transients) through refinement of parameters for LAT-detected GRBs, detection and characterization of GRBs not detected onboard, search and analysis of delayed high-energy afterglow emission; 2) flux history monitoring based on a maximum-likelihood method (\texttt{gtlike} science tool) for a predefined list of sources with subsequent addictions of publicly announced sources (flaring blazars, transient sources announced in ATels); 3) blind source detection on all-sky photon counts maps accumulated in 6-hour, 1-day, and 1-week intervals, through a fast sky map scan method based on two-dimensional Mexican Hat wavelet transform, thresholding and sliding cell algorithms \citep{ciprini07}; 4) transient and flare identification based on a variability test; 5) interactive LAT source catalogs; 6) multi-mission/multifrequency tools and archives (e.g. the error circle explorer and spectral energy distribution builder) linked to ASP and provided by the ASDC.

The list of discoveries triggered by FA-GSWs is substantial.  Some examples: 1) Flares from $\gamma$-ray blazars (extraordinary outbursts of 3C 454.3, very bright and large flares from other flat spectrum radio quasars like PKS 1510-089, 4C 21.35, AO 0235+164, PKS 1502+106, 3C 279, 3C 273, PKS 1622-253, PKS 1424-41. 2) Intense flares from BL Lac objects like Mkn 421, BL Lacertae, S5 0716+71. 3) Short/long duty cycles of other bright $\gamma$-ray  blazars. 4) Flares from gravitationally lensed $\gamma$-ray blazars like PKS 1830-211 and S3 0218+35. 5) Flares from radiogalaxies like NGC 1275 or NLSy1 like PMN J0948+0022. 6) Unidentified transients near the Galactic plane (e.g. J0910-5041, J0109+6134, Galactic center region). 6) Flares associated with high confidence with Galactic sources (such as the Crab nebula, nova V407 Cyg, microquasar Cyg X-3, binary system 1FGL J1018.6-5856, and possible flares associated with nova Mon 2012, and nova Sco 2012). 7) Intense $\gamma$-ray emission from the quiet and active sun (M/X class solar flares).

%
\section{Conclusions}
%

The all-sky variability monitor service run by the FA-GSW group represents the liaison between the \fermi LAT Collaboration and the multifrequency astrophysical and astroparticle communities \citep{thompson09,thompson11}. The role and activity of the FA-GSW shifter can be considered twofold. 1) \textit{Gamma-ray Flare Advocate task}: flaring sources approaching a daily photon flux of $10^{-6}$ photons (E$>$ 100 MeV) cm$^{-2}$ s$^{-1}$ deserve special attention (detection, localization, flux, photon index to be checked, photon counts maps and exposure maps to be reviewed).  Internal and public notes, ATels, Target of Opportunity (ToO) are submitted, and MW observing campaigns are organized when needed. 2) \textit{Gamma-ray Sky Watcher task}: results from the ASP pipeline in one-day and six-hour time intervals are checked, searching for transients, flaring sources, peculiar increasing or decreasing brightness trends, and  appearance of new candidate $\gamma$-ray sources, looking at positional multifrequency associations.

The FA-GSW activity has allowed us to find new $\gamma$-ray sources before \fermi catalog releases, identify sources that appeared  variable on short timescales,  discover a number of bright $\gamma$-ray flares and outbursts emitted by blazars and other AGN,  recognize transients from Galactic sources and low Galactic latitude sources not always associated with clear radio/optical/X-ray counterparts, and highlight the emission of the quiet and flaring Sun. In addiction FA-GSWs on duty carried out more than 40 ToOs with the \textit{Swift} satellite, showing the ideal synergy between the \fermi and \textit{Swift} missions.  FA-GSWs involved with the radio astronomy community have also organized MW observing campaigns targeted on single blazar sources and Galactic sources.

After 4.5 years of observations the output of \fermi ATels continues to increase steadily, showing that the $\gamma$-ray sky is showing interesting new features and new flaring sources.

%
\begin{acknowledgments}
%
\footnotesize{The \textit{Fermi} LAT Collaboration acknowledges generous ongoing support
from a number of agencies and institutes that have supported both the
development and the operation of the LAT as well as scientific data analysis.
These include the National Aeronautics and Space Administration and the
Department of Energy in the United States, the Commissariat \`a l'Energie Atomique
and the Centre National de la Recherche Scientifique / Institut National de Physique
Nucl\'eaire et de Physique des Particules in France, the Agenzia Spaziale Italiana
and the Istituto Nazionale di Fisica Nucleare in Italy, the Ministry of Education,
Culture, Sports, Science and Technology (MEXT), High Energy Accelerator Research
Organization (KEK) and Japan Aerospace Exploration Agency (JAXA) in Japan, and
the K.~A.~Wallenberg Foundation, the Swedish Research Council and the
Swedish National Space Board in Sweden.
\par Additional support for science analysis during the operations phase is gratefully
acknowledged from the Istituto Nazionale di Astrofisica in Italy and the Centre National d'\'Etudes Spatiales in France.
\par The FA-GSW coordinators Stefano Ciprini and Dario Gasparrini, and the LAT Collaboration express their gratitude to all the shifters that have served for the FA-GSW service until now, and express their gratitude to contributors for quick-look analysis software and scripting built for this service. FA-GSWs coordinators, shifters and the LAT Collaboration acknowledge people who worked to build and maintain L1 processing and L2 quick-look science analysis pipelines, like the Automatic Science Processing at the LAT Instrument Science Operation Center (ISOC), SLAC National Accelerator Laboratory, USA. The FA-GSW service is based also on \fermi\ services and products provided by the \fermi\ Science Support Center (FSSC), a constituent of the High Energy Astrophysics Science Archive Research Center (HEASARC) within the Astrophysics Science Division at the Goddard Space Flight Center, USA.
The FA-GSW service is based also on multifrequency on-line services and archives provided by the ASI Science Dara Center, Italy.
}
\end{acknowledgments}



\end{document}